\documentclass{WileyMSP-template}

\usepackage{hyperref}

\usepackage{amsmath}
\usepackage{amsfonts}
\usepackage{graphicx}
\usepackage{dcolumn}
\usepackage{bm}
\usepackage{dblfloatfix}

\usepackage[skip=2pt,font=footnotesize]{caption}
\usepackage[section]{placeins}
\usepackage{accents}

\newcommand{\rmi}[1]{\mathrm{i}{#1}}

\usepackage{xcolor}

\begin{document}

\pagestyle{fancy}
\rhead{\includegraphics[width=2.5cm]{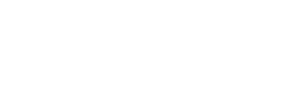}}

\title{Generalized Drude-Lorentz Model Complying with the Singularity Expansion Method}

\maketitle

\author{Isam Ben Soltane*}
\author{Félice Dierick}
\author{Brian Stout}
\author{Nicolas Bonod*}

\begin{affiliations}
Isam Ben Soltane\\
Aix Marseille Univ, CNRS, Centrale Marseille, Institut Fresnel, 13013 Marseille, France\\
Email Address: isam.ben-soltane@fresnel.fr\\

Félice Dierick\\
Aix Marseille Univ, CNRS, Centrale Marseille, Institut Fresnel, 13013 Marseille, France\\

Brian Stout\\
Aix Marseille Univ, CNRS, Centrale Marseille, Institut Fresnel, 13013 Marseille, France\\

Nicolas Bonod\\
Aix Marseille Univ, CNRS, Centrale Marseille, Institut Fresnel, 13013 Marseille, France\\
Email Address: nicolas.bonod@fresnel.fr\\
\end{affiliations}


\begin{abstract} 
Deriving analytical expressions of dielectric permittivities is required for numerical and physical modeling of optical systems and the soar of non-hermitian photonics motivates their prolongation in the complex plane. Analytical models are based on the association of microscopic models to describe macroscopic effects. However, the question is to know whether the resulting Debye Drude Lorentz models are not too restrictive. Here we show that the permittivity must be treated as a meromorphic transfer function that complies with the requirements of complex analysis. This function can be naturally expanded on a set of complex singularities. This singularity expansion of the dielectric permittivity allows us to derive a generalized expression of the Debye Drude Lorentz model that complies with the requirements of complex analysis and the constraints of physical systems. We show that the complex singularities and other parameters of this generalized expression can be retrieved from experimental data acquired along the real frequency axis. The accuracy of this expression is assessed for a wide range of materials including metals, 2D materials and dielectrics, and we show how the distribution of the retrieved poles helps in characterizing the materials.
\end{abstract}

\section{Introduction}

The description of the permittivity of materials \textit{via} analytical expressions is of fundamental interest in the field of optics and electromagnetism. Analytical models are relevant to provide simple and relevant descriptions of the interaction between materials and excitation fields~\cite{andreoli2021,shim2021}. They are also of strong interest for computational modeling and numerical methods such as time-domain numerical methods~\cite{kunz1993,taflove1995}. Besides the need of analytical description of the dielectric permittivity on real frequencies, modal analysis of open systems and non-hermitian photonics requires the description of the dielectric permittivities on the complex frequency plane~\cite{park2020symmetry,song2021plasmonic,colom2023crossing,wang2023non,Ferise2023Exceptional}. 
Common tools or models rely on combinations of the Drude, Lorentz and sometimes Debye models to describe the permittivity on spectral windows extending over a few dozens of wavelengths~\cite{oughstun2003,vial2005,vial2008,gharbi2020}. 

Alternatively, one can derive an expression of the permittivity (or the susceptibility) as a singularity expansion~\cite{grigoriev2014,garcia-vergara2017}. When real-valued fields are considered in the temporal domain, this becomes equivalent to the complex-conjugate pole-residue pairs model (ccprp)~\cite{han2006,han2010,sehmi2017,warner2022}, which approximates the permittivity using a finite set of complex conjugated pairs of poles contained within or close to a spectral window of interest, with the aim of providing simple analytical expressions to numerical methods~\cite{han2006,lin2012,prokopidis2018}. 

By expressing the permittivity as the transfer function of a physical system, we generalize these approaches and show that the singularity expansion method (SEM)~\cite{baum1971,bensoltane2023mosem} is the natural way to obtain an exact expression of the permittivity. The SEM can be recast into a form including the Drude, Lorentz and Debye models which we refer to as the Generalized Drude-Lorentz model (GDL). We discuss in particular what the additional terms appearing in the generalized Lorentz model translates into when we revert back to the temporal domain.  We propose an approach to retrieve the parameters of the GDL model. It is an optimization method relying on the auto-differentiation tools from current machine-learning libraries, and more specifically PyTorch~\cite{paszke2017}. We test this approach and show its accuracy with experimental data corresponding to the dielectric permittivity of $9$ materials including metals, dielectrics and 2D materials. Finally, we show how the distribution of the poles associated with Debye, Drude or Lorentz terms can characterize the behaviour of the material at different frequencies. While we focus on non-magnetic, isotropic media for simplicity but also and mainly because these account for most of the regularly encountered media, the study presented in this work can also be applied to non-isotropic media with tensors for both the permittivity and the permeability.

\section{Singularity expansion of the permittivity}

\subsection{Drude-Lorentz model}

The dielectric permittivity is often expressed as a combination of the Lorentz, Debye and Drude models. The Debye model being a special case of the Drude model, we will focus on Drude and Lorentz models. 
The Lorentz model is often used to model the behaviour of the electric charges of a medium excited by an electromagnetic-field. It leads to an expression of the permittivity as a sum of $N_e$ Lorentz functions, each associated with a group of electric charges possessing a different behaviour. The Drude model describes free charges in a metal or a gaz of charged particules. By simply putting them together, we obtain the classical Drude-Lorentz (DL) model:
\begin{eqnarray}
    \begin{aligned}
        \mathcal{E}(\omega) &= \varepsilon_0\left( \mathcal{E}_{\infty} - \frac{\omega_b^2}{\omega^2 + \rmi\omega\gamma} - \sum_{m=1}^{N_e} \frac{\omega_{p,m}^2}{(\omega^2 - \omega_{0,m}^2) + \rmi\omega\Gamma_m} f_m \right)
    \end{aligned}
    \label{eq:Drude_Lorentz_Model}
\end{eqnarray}
The Drude term is characterized by its plasma frequency $\omega_b$ and $\gamma$ which corresponds to a friction-like term acting on the free charged particles within the medium. The $N_e$ Lorentz terms have a similar expression, but are associated with forces acting upon bound charged particles. They mainly differ by the presence of the frequency $\omega_{0,m}$ which is associated with a spring-like force restoring the particles to their equilibrium position. $N_e$ usually ranging from $1$ and $5$, with the usual convention that $\sum_{m=1}^{N_e} f_m = 1$, which corresponds to $N_e$ groups of charged particles with different behaviours~\cite{jackson1999}

\subsection{The dielectric permittivity as a transfer function}

\begin{figure}[h!]
    \begin{center}
        \includegraphics[width=0.4\textwidth]{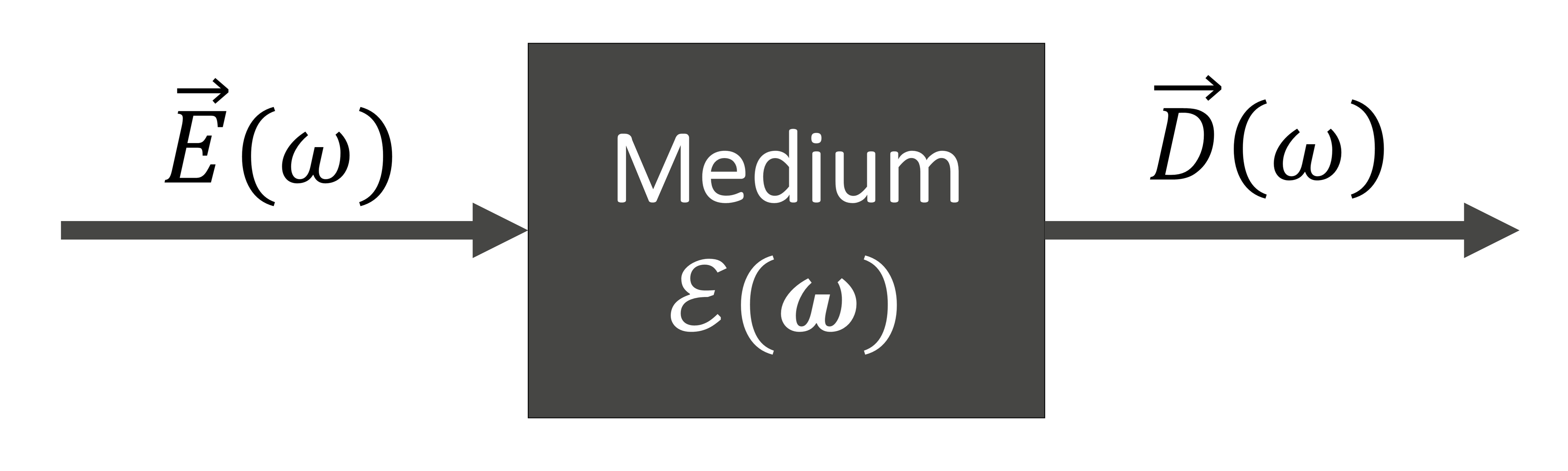}
    \end{center}
    \captionsetup{justification=centering}
    \caption{The linear dielectric permittivity of a material is defined in the constitutive relations by a transfer function linking the output displacement field $\vec{D}$ to the input electric field $\vec{E}$. It shares the properties of a linear transfer and can be analytically expressed using the singularity expansion method.}
    \label{Fig:fig1} 
\end{figure}

The permittivity $\mathcal{E}(\omega)$ is defined by the constitutive equation linking the electric field $\vec{E}$ to the displacement field $\vec{D}$. It has the form of a tensor and can be reduced to a scalar in isotropic media. The elements of this tensor link components of the exciting electric field to the components of the displacement field. Each component of the dielectric tensor behaves as a transfer function of a physical system. Transfer functions of linear and time invariant systems permit to retrieve the output field, here the displacement field $\vec{D}$, when the physical material is excited by an incoming field, here the electric field $\vec{E}$ (see Figure~(\ref{Fig:fig1})). In this study, we will consider isotropic, non-magnetic, linear and time invariant optical materials for which the dielectric permittivity reduces to a scalar that is obtained by taking the square of the refractive index $n(\omega)$.

The Singularity Expansion Method has been developed to expand the transfer functions in terms of their complex singularities~\cite{baum1971,grigoriev2013optimization,grigoriev2014,colom2018,bensoltane2023slab,bensoltane2023mosem}. Under the assumptions described in Equations (S3-S7) of the SI, it yields:
\begin{eqnarray}
    \begin{aligned}
        \mathcal{E}(\omega) &\approx \mathcal{E}_{\text{NR}} + \sum_{\ell=1}^{M} \frac{r_{I}^{(\ell)}}{\omega-\omega_I^{(\ell)}} + \sum_{\ell=1}^{N} \left[ \frac{r_p^{(\ell)}}{\omega-\omega_p^{(\ell)}} - \frac{\overline{r_p}^{(\ell)}}{\omega+\overline{\omega_p}^{(\ell)}} \right]
    \end{aligned}
    \label{SEM}
\end{eqnarray}
where $r_p^{(\ell)}$ (resp. $r_I^{(\ell)}$) is the residue of $\mathcal{E}$ at the complex singularity or pole $\omega_p^{(\ell)}$ (resp. imaginary pole $\omega_I^{(\ell)}$), and $\mathcal{E}_{\text{NR}}$ is the non-resonant term, a constant with a known expression depending on all the poles. Let us point out that, due to the inherent Hermitian symmetry, \textit{i.e.} $\overline{H(\omega)}=H(-\overline{\omega})$ of physical system resulting from the consideration of real-valued temporal signals~\cite{Nussenzveig1972}, the complex poles come in pairs  ($\omega_p^{(\ell)}$, $-\overline{\omega_p}^{(\ell)}$) and so do their associated residues ($r_p^{(\ell)}$, $-\overline{r_p}^{(\ell)}$)~\cite{han2006,sehmi2017,bensoltane2023mosem}.

\subsection{Generalized Drude-Lorentz expression}

We now aim at demonstrating that the singularity expansion in Equation~(\ref{SEM}) can be cast into a form that encompasses the Drude-Lorentz expression in Equation~(\ref{eq:Drude_Lorentz_Model}). For that purpose, we separate three different contributions in Equation~(\ref{SEM}) regarding the position of the poles in the complex plane. In agreement with the final value theorem, we impose a pole at the origin (the detailed justification can be found in the SI), and we separate the complex pairs of poles from the poles on the imaginary axis, which account for the contribution of (almost) free charges in the media. The purely imaginary poles, including the one at the origin, are associated with terms equivalent to the usual Drude or Debye relaxation terms as will be shown later. Let us call $\hat{\mathcal{E}}$ the expansion of $\mathcal{E}$:
\begin{eqnarray}
    \begin{aligned}
        \hat{\mathcal{E}}(\omega) &= \mathcal{E}_{\text{NR}} + \varepsilon_0 \left( \frac{r_0}{\omega} + \sum_{\ell=1}^{M} \frac{r_{I}^{(\ell)}}{\omega-\omega_I^{(\ell)}} + \sum_{\ell=1}^{N} \left[ \frac{r_p^{(\ell)}}{\omega-\omega_p^{(\ell)}} - \frac{\overline{r_p}^{(\ell)}}{\omega+\overline{\omega_p}^{(\ell)}} \right] \right) \\
    \end{aligned}
    \label{eq:truncated_MOSEM}
\end{eqnarray}
where $r_0$ is the residue of $\mathcal{E}$ at $0$. Let us stress that Equation~(\ref{eq:truncated_MOSEM}) would be the exact expression of the permittivity $\mathcal{E}$ if the expansion were not truncated for numerical reasons.

The imaginary poles in the truncated singularity expansion, including $0$, can be written as Drude terms. Let us consider the couple $(\omega_I^{(\ell)}, r_{I}^{(\ell)})$ of an imaginary pole and its associated imaginary residue. Setting $\gamma_l = -\text{Im}(\omega_I^{(\ell)})$ and $\omega_{b,\ell}\in\mathbb{C}$ such that $-\rmi\omega_{b,\ell}^2 / \gamma_\ell = r_{I}^{(\ell)}$ leads to:
\begin{eqnarray}
    \begin{aligned}
        \frac{r_{I}^{(\ell)}}{\omega-\omega_I^{(\ell)}} = -\frac{\omega_{b,\ell}^2}{\rmi\gamma_\ell}\left[ \frac{1}{\omega} - \frac{1}{\omega+\rmi\gamma_\ell} \right] + \frac{\omega_{b,\ell}^2}{\rmi\gamma_\ell} \frac{1}{\omega}
    \label{eq:complex_poles_GDL}
    \end{aligned}
\end{eqnarray}
Finally, we give an expression of $\gamma_0$, the residue of the pole $0$, taking into account $r_0$ and the residues of the other imaginary poles:
\begin{eqnarray}
    \begin{aligned}
        \gamma_0 = r_0 - \sum_{\ell=1}^{M}\frac{\omega_{b,\ell}^2}{\rmi\gamma_\ell}
    \end{aligned}
\end{eqnarray}
Thus, all the terms associated with the imaginary poles can be written as:
\begin{eqnarray}
    \begin{aligned}
        &\frac{r_0}{\omega} + \sum_{\ell=1}^{M} \frac{r_{I}^{(\ell)}}{\omega-\omega_I^{(\ell)}} = \frac{\gamma_0}{w} - \sum_{\ell=1}^{M} \frac{\omega_{b,\ell}^2}{\omega^2 + \rmi\omega\gamma_\ell}
    \end{aligned}
    \label{eq:imaginary_poles_GDL}
\end{eqnarray}
The sum over $\ell$ is equivalent to considering $M$ groups of freely moving charges, each associated with different parameters $\gamma_\ell$ and $\omega_{b,\ell}$. Since $\omega_{b,\ell}^2$ can be negative, Equation~(\ref{eq:imaginary_poles_GDL}) spans a larger set of functions than the regular Drude model written in Equation~(\ref{eq:Drude_Lorentz_Model}) and for which $\omega_{b,\ell}$ is a positive frequency.

Similarly, the terms involving the complex poles $\omega_p^{(\ell)}$ can be written as Lorentz-like terms. For any pole $\omega_p^{(\ell)}$, we have:
\begin{eqnarray}
    \begin{aligned}
        &\frac{r_p^{(\ell)}}{\omega-\omega_p^{(\ell)}} - \frac{\overline{r_p}^{(\ell)}}{\omega+\overline{\omega_p}^{(\ell)}} = 2\frac{\rmi\omega\text{Im}(r_p^{(\ell)}) + \text{Re}(r_p^{(\ell)}~\overline{\omega_p}^{(\ell)})}{\omega^2 - |\omega_p^{(\ell)}|^2 - 2i\omega\text{Im}~(\omega_p^{(\ell)})}
    \end{aligned}
\end{eqnarray}
Let us set $\omega_{0,\ell}=|\omega_p^{(\ell)}|$ and $\Gamma_\ell=-2\text{Im}(\omega_p^{(\ell)})$, and two constants $s_{1,\ell}$ and $s_{2,\ell}$ such that $s_{1,\ell} \Gamma_l = -2\text{Im}(r_p^{(\ell)})$ and $s_{2,\ell} \omega_{0,\ell}^2 = -2\text{Re}(r_p^{(\ell)}~\overline{\omega_p}^{(\ell)})$. The previous expression becomes:
\begin{eqnarray}
    \begin{aligned}
        \frac{r_p^{(\ell)}}{\omega-\omega_p^{(\ell)}} - \frac{\overline{r_p}^{(\ell)}}{\omega+\overline{\omega_p}^{(\ell)}} &= -\frac{\rmi s_{1,\ell}\omega\Gamma_\ell + s_{2,\ell}\omega_{0,\ell}^2}{(\omega^2 - \omega_{0,\ell}^2) + \rmi\omega\Gamma_\ell} = \frac{P_{(\ell)}(\omega)}{Q_{(\ell)}(\omega)}
    \end{aligned}
    \label{eq:generalized_lorentz}
\end{eqnarray}
to provide the Generalized Drude Lorentz expression:
\begin{eqnarray}
    \begin{aligned}
        \hat{\mathcal{E}}(\omega) &=  \mathcal{E}_{\text{NR}} + \varepsilon_0 \left( \frac{\gamma_0}{w} - \sum_{\ell=1}^{M} \frac{\omega_{b,\ell}^2}{\omega^2 + \rmi\omega\gamma_\ell} \right.\\
        &\left. - \sum_{\ell=1}^{N} \frac{\rmi s_{1,\ell}\omega\Gamma_\ell + s_{2,\ell}\omega_{0,\ell}^2}{(\omega^2 - \omega_{0,\ell}^2) + \rmi\omega\Gamma_\ell} \right)
    \end{aligned}
    \label{eq:GDL_pre_pytorch}
\end{eqnarray}
The $N$ generalized Lorentz function in Equation~(\ref{eq:GDL_pre_pytorch}) differ from the classical Lorentz functions in Equation~(\ref{eq:Drude_Lorentz_Model}) by the frequency-dependent imaginary terms $\rmi s_{1,\ell}\omega\Gamma_\ell$ in the numerator $P_{(\ell)}$ which are tantamount to non-real valued residues associated with the Lorentz poles $r_p^{(\ell)}$. 
Let us look at the contribution of only one generalized Lorentz function $P_{(\ell)} / Q_{(\ell)}$ to the permittivity. The relationship between the displacement field $\vec{D}(\omega)$ and the electric field $\vec{E}(\omega)$ now reads as:
\begin{eqnarray}
    \begin{aligned}
        Q_{(\ell)}(\omega)\vec{D}(\omega) = \varepsilon_0\ P_{(\ell)}(\omega) \vec{E}(\omega)
    \end{aligned}
    \label{eq:generalized_lorentz_sep}
\end{eqnarray}
Or, by replacing $P_{(\ell)}$ and $Q_{(\ell)}$ by their expressions:
\begin{eqnarray}
    \begin{aligned}
        &-\left((\omega^2 - \omega_{0,\ell}^2) + \rmi\omega\Gamma_\ell\right) \vec{D}(\omega) = \varepsilon_0\ \left(\rmi s_{1,\ell}\omega\Gamma_\ell + s_{2,\ell}\omega_{0,\ell}^2\right) \vec{E}(\omega)
    \end{aligned}
    \label{eq:generalized_lorentz_sep_full}
\end{eqnarray}
It follows that the additional imaginary term  $\rmi\omega\Gamma_\ell$ accounts for a new contribution of the electric field  to the displacement field when we compare it to the classical Lorentz function numerator $s_{2,\ell}\omega_{0,\ell}^2$. We can perform an inverse Laplace transform of Equation~(\ref{eq:generalized_lorentz_sep_full}):
\begin{eqnarray}
    \begin{aligned}
        &\left( (\partial_t^2 +  \omega_{0,\ell}^2) + \Gamma_\ell\partial_t \right)[\vec{d}](t) = \varepsilon_0\left( s_{1,\ell}\Gamma_\ell\partial_t - s_{2,\ell}\omega_{0,\ell}^2 \right)[\vec{e}](t)
    \end{aligned}
    \label{eq:temporal_generalized_lorentz}
\end{eqnarray}
In the temporal domain, the imaginary term translates into a contribution of the scaled derivative of the electric field $s_{1,\ell}\Gamma_\ell\partial_t[\vec{e}](t)$ to the displacement field $\vec{d}(t)$. This new dependency can be explained \textit{via} the constitutive relation linking the displacement field, the electric field, and the polarization density field $\vec{p}$: $\vec{d} = \varepsilon_0 \vec{e} + \vec{p}$. The polarization density vector is connected to the electric field \textit{via} the electric susceptibility. As stated in reference~\cite{garcia-vergara2017}, this is equivalent to considering two linear operators $\mathcal{L}_1$ and $\mathcal{L}_2$ such that $\mathcal{L}_1\left[\vec{p}\right](t) = \varepsilon_0\mathcal{L}_2\left[\vec{e}\right](t)$, which results in $mathcal{L}_1[\vec{d}](t) = \varepsilon_0\left(\mathcal{L}_1 + \mathcal{L}_2\right) \left[\vec{e}\right](t)$. 
$\mathcal{L}_1$ and $\mathcal{L}_2$ can be retrieved \textit{via} identification in Equation~(\ref{eq:temporal_generalized_lorentz}):
\begin{eqnarray}
    \begin{aligned}
        \mathcal{L}_1 &= \left(\partial_t^2 +  \omega_{0,\ell}^2\right) + \Gamma_\ell\partial_t \\
        \left(\mathcal{L}_1 + \mathcal{L}_2\right) &= s_{1,\ell}\Gamma_\ell\partial_t - s_{2,\ell}\omega_{0,\ell}^2
    \end{aligned}
\end{eqnarray}
In the classical Lorentz approach, the last term reduces to $\left(\mathcal{L}_1 + \mathcal{L}_2\right) = - s_{2,\ell}\omega_{0,\ell}^2$.

Let us combine Equation~(\ref{eq:complex_poles_GDL}) and Equation~(\ref{eq:imaginary_poles_GDL}) to obtain the following alternative form of the truncated SEM expression:
\begin{eqnarray}
    \begin{aligned}
        \hat{\mathcal{E}}(\omega) &=  \mathcal{E}_{\text{NR}} + \varepsilon_0 \left(  \frac{\gamma_0}{\omega} - \sum_{\ell=1}^{M} \frac{\omega_{b,\ell}^2}{\omega^2 + \rmi\omega\gamma_\ell} - \sum_{\ell=1}^{N} \frac{\rmi s_{1,\ell}\omega\Gamma_\ell + s_{2,\ell}\omega_{0,\ell}^2}{(\omega^2 - \omega_{0,\ell}^2) + \rmi\omega\Gamma_\ell} \right)
    \end{aligned}
    \label{eq:GDL}
\end{eqnarray}
Equation~(\ref{eq:GDL}) is what we refer to as the Generalized Drude-Lorentz (GDL) model as it encompasses both models (and also the Debye model) and generalizes the Lorentz terms \textit{via} the additional imaginary terms.It follows that the permittivity can be equivalently approximated equivalently as a pole expansion with Equation~(\ref{eq:truncated_MOSEM}), and as a GDL expression with Equation~(\ref{eq:GDL}).

\section{Retrieving the parameters from experimental data}

We show how to retrieve the parameters of the GDL model using experimental data at real frequencies. This approach is tested on the $9$ materials listed in Table S1 of the SI, which include oxides, metals and 2D materials. 

\subsection{Analytical continuation of the permittivity from experimental data}

Each material is associated with a set of experimental data points $(\omega_i, \varepsilon_i)$ with $\varepsilon_i=\mathcal{E}(\omega_i)$ the permittivity measured at the real frequency $\omega_i$. Starting from the expression of the GDL model in Equation~(\ref{eq:GDL}), we wish to find the set of parameters $\mathcal{P}=\{\mathcal{E}_{NR}, \gamma_0, \omega_{b,\ell}, \gamma_\ell, s_{1,\ell}, \Gamma_\ell, s_{2,\ell}, \omega_{0,\ell}\}$, which analytically expands the permittivity into the complex frequency plane while minimizing the distance between the experimental points $\varepsilon_i$ and the estimated GDL value $\tilde{\varepsilon}_{i,\mathcal{P}}=\tilde{\mathcal{E}}(\omega_i, \mathcal{P})$. The distance between the model and experimental data is defined by error or loss function $L$:
\begin{eqnarray}
    \begin{aligned}
        \mathcal{P}^* = \underset{\mathcal{P}}{\arg\min}~L(\varepsilon_i, \tilde{\varepsilon}_{i,\mathcal{P}})
    \end{aligned}
    \label{eq:opt}
\end{eqnarray}
The final set of parameters $\mathcal{P}^*$ is obtained with auto-differentiation, a method that has been increasingly used in optimization problems in electromagnetism thanks to the versatility and efficiency that this tool offers~\cite{minkov2020,alagappan2023,so2022design,so2022holography}. In this study, we use the open-source tools provided by the machine-learning library PyTorch. The loss function $L$, and the optimization process are described in Equations~(S8-S18) of the SI. 
\begin{figure}[h!]
    \begin{center}
        \includegraphics[width=\textwidth]{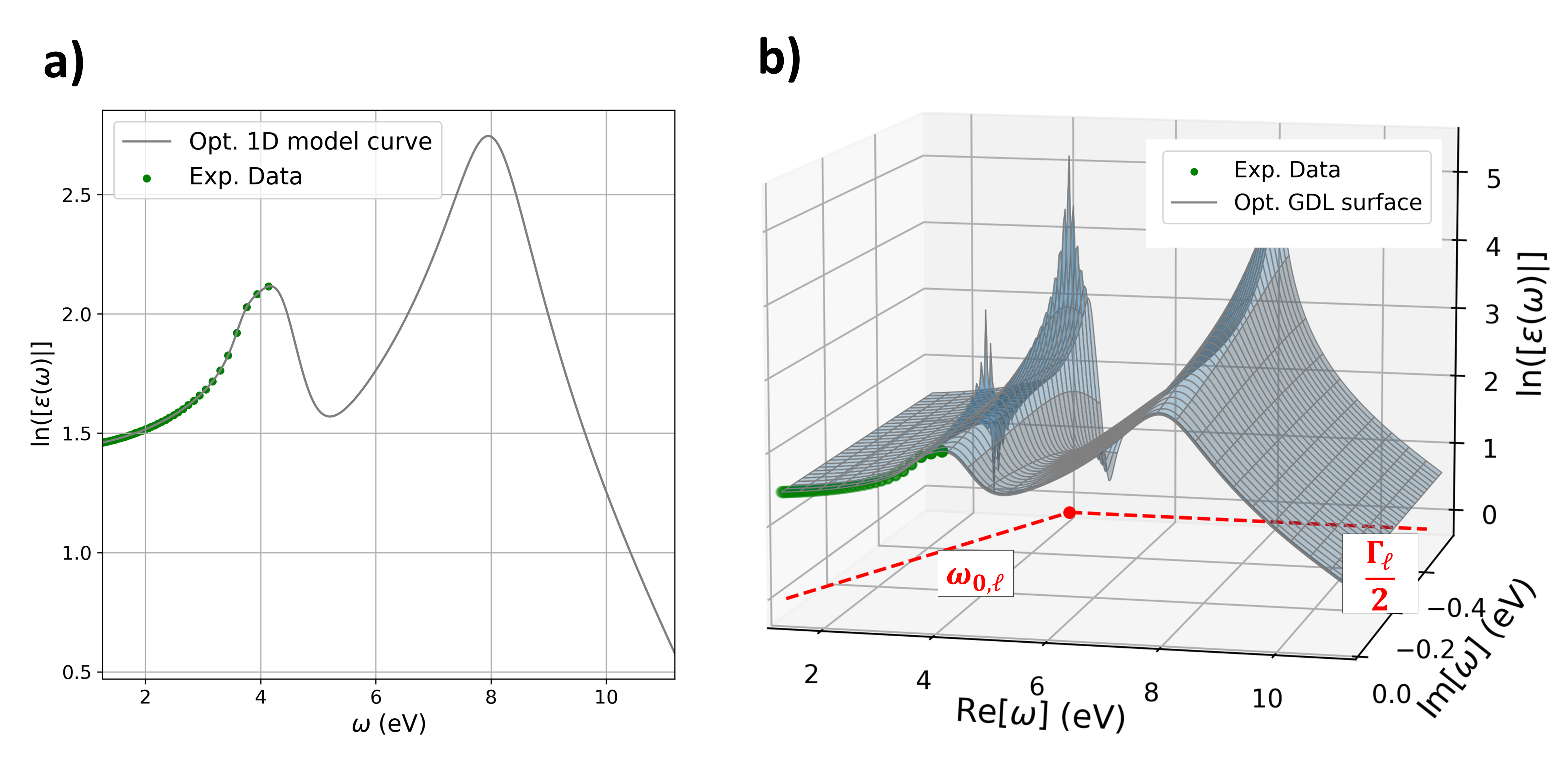}
    \end{center}
    \captionsetup{justification=centering}
    \caption{Optimization of a model for the dielectric permittivity of $\text{TiO}_2$ using experimental data. (a) Classical $1D$ optimization by a curve that fits experimental data available at low frequencies (below $5$ eV). The $1D$ model provides a value of the dielectric permittivity at any real frequency. (b) Retrieval of the GDL parameters by auto-differentiation based on gradient-descent-like optimization. The parameters are chosen to minimize a loss function $L$ as described in Equation~(\ref{eq:opt}). The process is equivalent to the optimization of a surface in the complex frequency plane by moving the poles $\omega_p^{(\ell)}$ to fit the experimental data available at real frequencies only.}
    \label{Fig:fig2} 
\end{figure}
The analytical continuation of the permittivity using experimental data is illustrated in Figure~\ref{Fig:fig2} in the case of $\text{TiO}_2$, where the calculated function is represented by a $1D$ curve at real frequencies in Figure~\ref{Fig:fig2} (a), and as a surface in the complex frequency plane in Figure~\ref{Fig:fig2} (b).
Let us stress that by setting the coefficients $s_{1,\ell}$ to $0$ in Equation~(\ref{eq:GDL}), we end up reducing the generalized Lorentz terms to the classical Lorentz terms. To justify the use of the GDL model rather than the classical DL model, the two of them are compared in Figures S1 and S2 of the SI, where we observe that the GDL model provides more accurate expressions than the DL model when a small number of singularities are involved. This is especially true in non-metallic media. As we add more and more singularities, the models both start to account for the noise rather than the dielectric permittivity data itself, which leaves no room for comparison.

\begin{figure}[h!]
    \begin{center}
        \includegraphics[width=\textwidth]{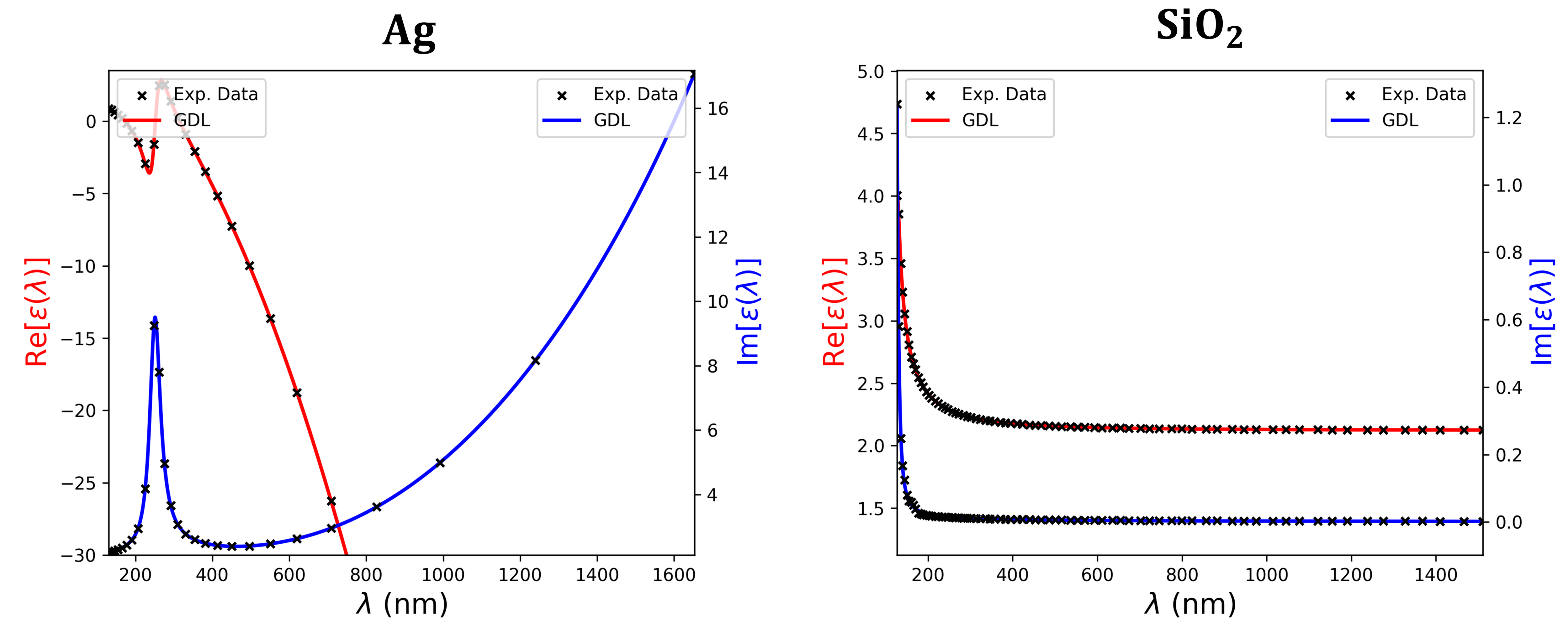}
    \end{center}
    \captionsetup{justification=centering}
    \caption{Real (red line, left axis) and imaginary (blue line, right axis) parts of the permittivity functions of $\text{Ag}$ and $\text{SiO}_2$ calculated with the optimized GDL model (full line) and compared to the experimental data (dotted line). The permittivity functions are expressed as functions of the wavelength instead of the frequency, to better highlight the spectral windows of the experimental data ranging from the UV to the NIR. The amplitude of the permittivity of $\text{Ag}$ quickly increases for wavelength above $380$nm, in agreement with its metallic behaviour. In the case of $\text{SiO}_2$, the permittivity is quasi-constant in the whole spectral window, with an imaginary part close to $0$ as one would expect in an absorbtion-free dieletric material.}
    \label{Fig:fig3} 
\end{figure}

We quantify the accuracy of the approach \textit{via} the calculation of the relative $L_2$ and $L_\infty$ errors between the experimental data and the optimized models (see Equations~(S8-S9) in the SI). The errors are summarized in Table~\ref{tab:results}, and show the excellent accuracy reached with the GDL model in a broad spectral range and for a large set of materials. 
\begin{table}[h!]
    \centering
    \begin{tabular}{ | m{0.17\textwidth}<{\centering} | m{0.12\textwidth}<{\centering} | m{0.12\textwidth}<{\centering} | m{0.22\textwidth}<{\centering} | m{0.22\textwidth}<{\centering} | } 
        \hline
        \multicolumn{5}{|c|}{Relative Errors of the Fitting} \\

        \hline
        Material & N & M & $\text{error}_2$ ($\%$) &  $\text{error}_\infty$ ($\%$)\\
        
        \hline
        $\text{Au}$ & 5 & 1 & 0.272 & 0.247\\
        
        \hline
        $\text{Ag}$ & 3 & 2 & 0.854 & 0.964\\
        
        \hline
        $\text{Co}$ & 5 & 1 & 0.145 & 0.070\\
        
        \hline
        $\text{Cr}$ & 3 & 1 & 0.542 & 0.834\\
        
        \hline
        $\text{HfO}_2$ & 2 & 0 & 0.074 & 0.276\\
        
        \hline
        $\text{SiO}_2$ & 2 & 0 & 0.287 & 1.192\\
        
        \hline
        $\text{TiO}_2$ & 3 & 0 & 0.237 & 0.763\\
        
        \hline
        $\text{Ta}_2\text{O}_5$ & 6 & 0 & 0.515 & 1.180\\
        
        \hline
        \text{Graphene} & 5 & 1 & 0.331 & 0.843\\
        
        \hline
    \end{tabular}
    \captionsetup{justification=centering}
    \caption{Relative $L_2$ error ($\text{error}_2$) and $L_\infty$ error ($\text{error}_\infty$) of the optimized GDL expressions of the permittivity for the $9$ materials listed in Table S1 of the SI. The lower the errors, the higher the accuracy. The parameters $N$ and $M$ represent respectively the number of generalized Lorentz terms and the number of Drude/Debye terms in Equation~(\ref{eq:GDL}).}
    \label{tab:results}
\end{table}
We observe an excellent agreement of the GDL model with the experimental data for all $9$ materials in a large spectral window, using a very small set of poles.  In addition, the GDL model outperforms the DL model due to the presence of the imaginary terms contained within the generalized Lorentz terms in Equation~(\ref{eq:generalized_lorentz}), which are required for the GDL model to comply with the SEM. Figures S1 and S2 of the SI support this observation by showing that the classical DL model struggles to reach the same accuracy with small sets of poles.
The applicability of this expansion for a wide range of materials comes from its compliance with complex analysis required by the fact that the dielectric permittivity is a linear transfer function. We graphically highlight these results for $\text{SiO}_2$ and $\textit{Ag}$ in Figure~\ref{Fig:fig3} where their associated GDL model are compared to the experimental data.

\subsection{Characterization of the materials with the distribution of the poles}

\begin{figure}[h!]
    \begin{center}
        \includegraphics[width=\textwidth]{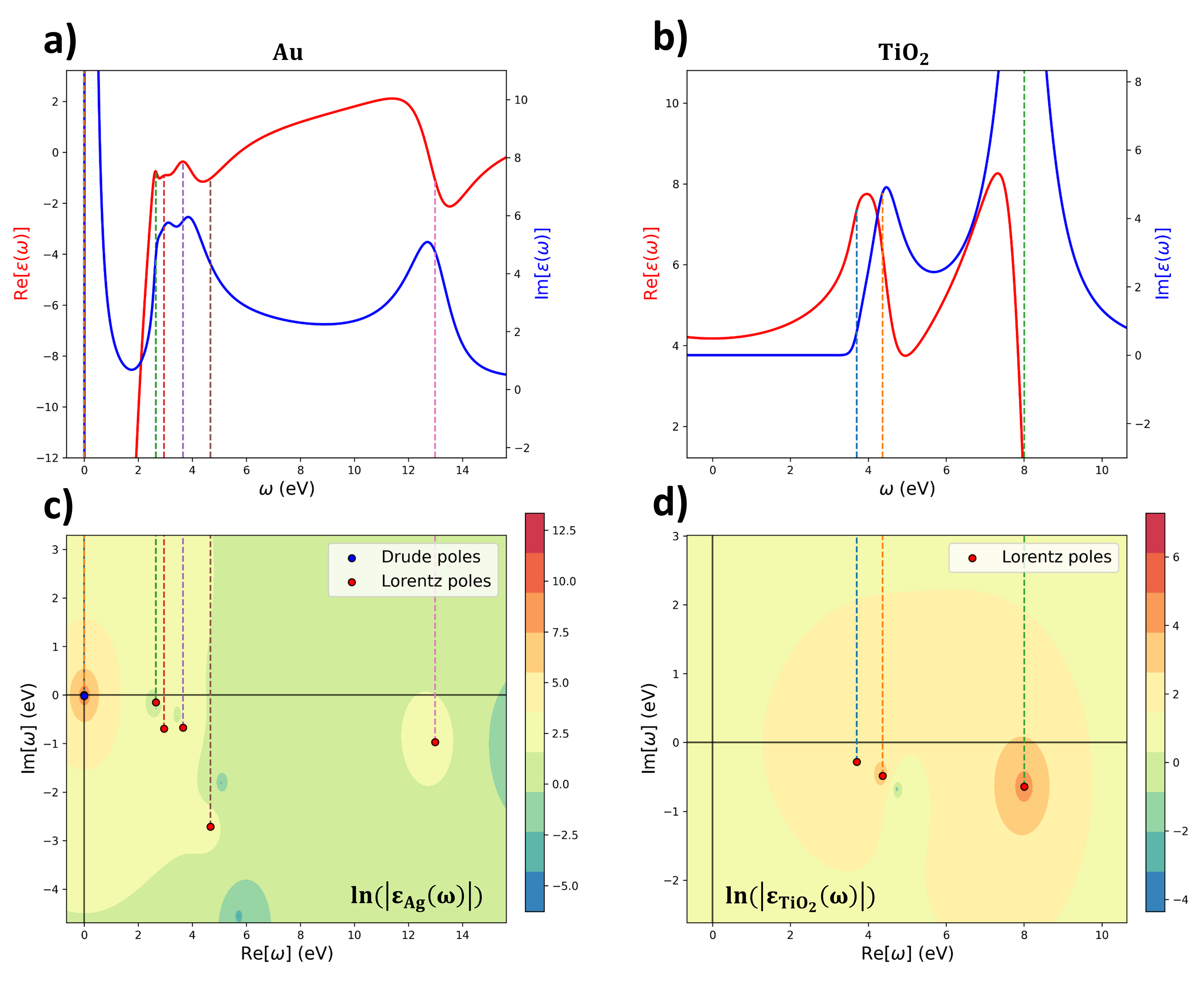}
    \end{center}
    \captionsetup{justification=centering}
    \caption{(a, b) Real and imaginary parts of the permittivity of $\text{Au}$ (a) and $\text{TiO}_2$ (b) at real frequencies, calculated using the optimized GDL. (c,d) Log-amplitudes of the dielectric permittivity function in the complex frequency plane obtained with the GDL model. The poles associated with Lorentz or Drude terms are indicated by red dots in the complex $\omega$-plane. Vertical dashed lines highlight the link between the real parts of the poles and the spectral features of the real and imaginary parts of the dielectric permittivity at real frequencies.}
    \label{Fig:fig4} 
\end{figure}

By inverting the relations introduced in Equations~(\ref{eq:imaginary_poles_GDL}) and (\ref{eq:complex_poles_GDL}), we convert the Debye, Drude and generalized Lorentz terms back into complex poles and residues which are then highlighted in the complex frequency plane. The poles and residues which do not contribute to significantly improving the matching between the model and the experimental data mostly help in fitting experimental or numerical noise and are therefore removed. The numbers of imaginary poles $M$ and pairs of poles $N$ are given in Table~\ref{tab:results}. We plot in Figure~\ref{Fig:fig4} the log-amplitude of the permittivity of $\text{Au}$ and $\text{TiO}_2$ in the complex frequency-plane and show qualitatively the correlation between the distribution of the poles and the amplitude of the permittivity at real frequencies.

At first glance, the real part of the poles is a good indicator of resonances and anti-resonances phenomena at real-frequencies. The distribution of the poles can be used to directly assess the behaviour of the material, \textit{i.e.} dielectric or metal, as one would expect. Let us first consider the case of gold ($\text{Au}$). At low frequencies, the free electrons of the metal allow for a great reflectivity, and thus a high amplitude of the permittivity, and in particular a real part far below $-1$. This observation is associated with the presence of Drude poles on the imaginary axis along with a singularity at the origin. After $2.7$ eV, which corresponds to approximately $430$nm, the material starts behaving like a dielectric with a rapidly varying and sign-changing real part. This phenomenon occurs at the real frequency of the first Lorentz pole. In low-absorption dielectrics such as $\text{TiO}_2$, the imaginary part of the permittivity is close to $0$, and the permittivity is quasi-constant on a wide spectral window. In this case, the Lorentz poles are located far from that window, and thus weakly affect them. Overall, the Lorentz poles mostly mark the transitions between the different regimes, \textit{i.e.} the metallic regime, the dispersive dielectric regime, and the near-constant regime, while the Drude poles are associated with extremely high values of the permittivity due to the free charged particles within a metal. Therefore, it is possible to characterize the nature of the materials based on the position the poles.

\begin{table}[h!]
	\centering
	\begin{tabular}{ | m{0.15\textwidth}<{\centering} | m{0.15\textwidth}<{\centering} | m{0.15\textwidth}<{\centering} | m{0.15\textwidth}<{\centering} | m{0.15\textwidth}<{\centering} | }
		 \hline
		 \multicolumn{5}{|c|}{ Au }\\
		 \hline
		\multicolumn{5}{|c|}{Drude/Debye terms}\\

		\hline
		$M$ & \multicolumn{2}{|c|}{$\omega_{b,\ell}$ (eV)} & \multicolumn{2}{|c|}{$\gamma_\ell$ (eV)}\\

		\hline
		 1  & \multicolumn{2}{|c|}{ 6.0e+00+0.0e+00j } & \multicolumn{2}{|c|}{ 1.8e-02 } \\

		\hline
		\multicolumn{5}{|c|}{Lorentz terms}\\

		\hline
		$N$ & $s_{1,\ell}$ & $\Gamma_\ell$ (eV) & $\omega_{0,\ell}$ (eV) & $s_{2,\ell}$\\

		\hline
		 1  &  -9.0e-01  &  3.0e-01  &  2.7e+00  &  1.4e-01 \\

		\hline
		 2  &  -2.1e+00  &  1.4e+00  &  3.0e+00  &  2.7e+00 \\

		\hline
		 3  &  -3.6e+00  &  1.3e+00  &  3.7e+00  &  6.9e-01 \\

		\hline
		 4  &  -7.9e-01  &  5.4e+00  &  5.4e+00  &  9.5e-02 \\

		\hline
		 5  &  2.3e+00  &  1.9e+00  &  1.3e+01  &  5.3e-01 \\

		\hline
		 \multicolumn{5}{|c|}{Non-resonant term}\\
		\hline
		$\mathcal{E}_{NR}$ & \multicolumn{4}{|c|}{ 4.7e-01 }\\

		\hline
	\end{tabular}
	\captionsetup{justification = centering}
	\caption{Parameters of the optimized GDL expression from Equation~\ref{eq:GDL} obtained \textit{via} auto-differentiation for the permittivity of $\text{Au}$. The frequencies $\omega_{0,\ell}$ are the moduli of the Lorentz poles, and are thus systematically larger than the imaginary parts $\Gamma_\ell$ of the poles. Since $\text{Au}$ is a metal, the coefficient $\gamma_0$ is null, and only the Drude terms remain instead of the Drude/Debye terms. $-\rmi\gamma_\ell$ are purely imaginary Drude poles, while the frequencies $\omega_{b,\ell}$ appear in the residues associated with the poles and are related to the frequency range within which the material still behaves as a metal.}
	\label{tab:Au_GDL}
\end{table}

The values of the poles put forth in Figure~\ref{Fig:fig4} for $\text{Au}$ and their associated residues are presented in Table~\ref{tab:Au_GDL}. The moduli of the Lorentz poles $\omega_{0,\ell}$ located in the studied frequency window (approximately $1$ to $10$ eV) are in concordance with the tabulated band gaps found in the literature~\cite{rangel2012}, which goes to show that the retrieved poles hold information regarding the behaviour of the materials at the microscopic scale and below. 

\section{Conclusion}

To conclude, the dielectric permittivity is a transfer function that is naturally described by its singularity expansion in the harmonic domain, which encompasses the ccprp method. We have shown that this expansion can be recast into a generalized Drude-Lorentz model, which fully complies with the requirements and constraints of physical systems. The generalized Lorentz model, in particular, contains an additional frequency-dependent imaginary term which corresponds to a contribution of the first derivative of the electric field to the displacement field in the temporal domain. This additional imaginary term is associated with a complex residue if the harmonic domain, which sets the generalized Lorentz terms apart from the classical Lorentz terms. 
We have proposed a method relying on auto-differentiation to retrieve the parameters of the generalized Drude-Lorentz model and have applied it to the retrieval of the permittivity of $9$ different materials including oxides, metals and 2D materials. We systematically obtained very low errors using a small set of Drude and generalized Lorentz terms in a wide spectral window extending from the UV to the NIR, showcasing not only the efficiency of the method, but also the high-accuracy of the generalized model. 
Finally, we have shown that, by converting back the parameters of the generalized Drude-Lorentz model into poles and residues, it is possible to characterize a medium \textit{via} the distribution of singularities of the permittivity. Drude singularities are associated with a singularity at the origin and purely imaginary poles and residues very close to the origin. In transparent medium, a near-constant permittivity over a frequency range translates into a lack of singularities into and close to the associated complex frequency window. The transition from one regime to another (for instance metallic to dieletric behaviour) comes along with a Lorentz pole with a relatively small imaginary part corresponding to a resonance phenomenon, hence a rapidly varying permittivity. Far from only adding degrees of freedom in the generalized Drude-Lorentz model, the imaginary parts of the residues associated with Lorentz poles are what allows the expression of the permittivity to be so accurate with so few of singularities.

\medskip
\textbf{Supporting Information} \par 
Supporting Information is available from the Wiley Online Library or from the author.

\medskip
\textbf{Acknowledgements} \par 
This work was funded by the French National Research Agency ANR
Project DILEMMA (ANR-20-CE09-0027). The authors thank Guillaume Demesy for the the fruitful  discussions and remarks.

\medskip
\bibliographystyle{MSP}
\bibliography{biblio}

\end{document}


\pagestyle{fancy}
\rhead{\includegraphics[width=2.5cm]{empty_logo.png}}

\title{Supporting Information for "Generalized Drude-Lorentz model complying with the singularity expansion of transfer functions"}

\maketitle

\author{Isam Ben Soltane*}
\author{Félice Dierick}
\author{Brian Stout}
\author{Nicolas Bonod}

\begin{affiliations}
Isam Ben Soltane\\
Aix Marseille Univ, CNRS, Centrale Marseille, Institut Fresnel, 13013 Marseille, France\\
Email Address: isam.ben-soltane@fresnel.fr\\

Félice Dierick\\
Aix Marseille Univ, CNRS, Centrale Marseille, Institut Fresnel, 13013 Marseille, France\\

Brian Stout\\
Aix Marseille Univ, CNRS, Centrale Marseille, Institut Fresnel, 13013 Marseille, France\\

Nicolas Bonod\\
Aix Marseille Univ, CNRS, Centrale Marseille, Institut Fresnel, 13013 Marseille, France\\
Email Address: nicolas.bonod@fresnel.fr\\
\end{affiliations}


\section{The permittivity as a transfer function}

We write the relation between the displacement field $\vec{D}$ and the electric field $\vec{E}$ in the harmonic domain, in a non-magnetic medium, using a Taylor expansion:
\begin{eqnarray}
    \begin{aligned}
        \frac{1}{\varepsilon_0}\vec{D}(\omega) = \vec{E}(\omega) + \bar{\bar{\chi}}^{(1)}(\omega)\vec{E}(\omega) + \bar{\bar{\chi}}^{(2)}(\omega)\vec{E}^2(\omega) + ...
    \end{aligned}
\end{eqnarray}
where $\bar{\bar{\chi}}^{(n-1)}$ is the $n^{\rm th}$ order electric susceptiblity tensor. Under the hypothesis that the tensors are all meromorphic functions of the frequency $\omega$, the descriptions that follow would still apply to all their components, leading to a model for non-linear permittivity. However, we work with the common assumption that the considered media are linear and isotropic with respect to the electric field. Thus:
\begin{eqnarray}
    \begin{aligned}
        \vec{D}(\omega) &= \varepsilon_0 (1 + \chi(\omega)) \vec{E}(\omega) \\
                        &= \mathcal{E}(\omega) \vec{E}(\omega)
    \end{aligned}
    \label{eq:rel_D_E}
\end{eqnarray}
with $\chi(\omega)$ and $\mathcal{E}(\omega)$ the scalar electric susceptibility and permittivity respectively. We choose to study the permittivity rather than the susceptiblity, but once again, what follows would also work with the latter. We see from Equation~(\ref{eq:rel_D_E}) that $\mathcal{E}$ is a transfer function. The singularity expansion method (SEM) can thus be applied\cite{baum1971, bensoltane2023mosem}. Assuming only poles of order $1$, it yields:
\begin{eqnarray}
    \begin{aligned}
        \mathcal{E}(\omega) &= \varepsilon_0 \left( \mathcal{E}(a) - \sum_{p} \frac{r_p}{a-\omega_p} + \sum_{p} \frac{r_p}{\omega-\omega_p} \right)
    \end{aligned}
    \label{eq:base_SEM}
\end{eqnarray}
with $r_p$ the residue of $\mathcal{E}$ at the simple pole $\omega_p$ and $a$ an arbitrary complex frequency. Although the set of singularities of $\mathcal{E}$ is potentially infinite, the SEM must be numerically truncated. We separate the resulting $M$ purely imaginary poles $\omega_I^{(l)}$ from the other $N$ complex ones $\omega_p^{(l)}$. In addition, the hermitian symmetry resulting from the real-valued fields considered in the temporal domain imposes that the complex poles and their associated residues come in pairs:
\begin{eqnarray}
    \begin{aligned}
        \mathcal{E}(\omega) &\approx \varepsilon_0 \left( \mathcal{E}(a) - \sum_{l=1}^{M} \frac{r_{I}^{(l)}}{a-\omega_I^{(l)}} - \sum_{l=1}^{N} \left[ \frac{r_p^{(l)}}{a-\omega_p^{(l)}} - \frac{\overline{r_p}^{(l)}}{a+\overline{\omega_p}^{(l)}} \right] \right. \\
        &+ \left. \sum_{l=1}^{M} \frac{r_{I}^{(l)}}{\omega-\omega_I^{(l)}} + \sum_{l=1}^{N} \left[ \frac{r_p^{(l)}}{\omega-\omega_p^{(l)}} - \frac{\overline{r_p}^{(l)}}{\omega+\overline{\omega_p}^{(l)}} \right] \right) \\
    \end{aligned}
    \label{eq:truncated_SEM_no_imag}
\end{eqnarray}
The Kramers-Kronig relations must hold for the permittivity, which is thus analytic in the upper half of the complex-frequency plane. This is equivalent to considering the media as passive, causal systems which only possess stable poles, \textit{i.e.} singularities with a negative imaginary part. Finally, we expect the displacement field to either fade or remain constant in the temporal domain after a sufficiently long time, no matter which non-diverging electric field excites the medium. The final value theorem applied to the permittivity thus reads as:
\begin{eqnarray}
    \begin{aligned}
        \lim_{t \rightarrow +\infty} \varepsilon(t) = \lim_{\omega \rightarrow 0+} \rmi\omega\mathcal{E}(\omega) = cst
    \end{aligned}
\end{eqnarray}
In case the displacement field does not completely disappear, the theorem only holds if $0$ is a pole of $\mathcal{E}$ of order $1$ at most. We therefore make the assumption that $0$ is always a singularity of $\mathcal{E}$ leading to the following singularity expansion of $\hat{\mathcal{E}}$:
\begin{eqnarray}
    \begin{aligned}
        \hat{\mathcal{E}}(\omega) &= \mathcal{E}_{NR} + \varepsilon_0 \left(  \frac{r_0}{\omega} + \sum_{l=1}^{M} \frac{r_{I}^{(l)}}{\omega-\omega_I^{(l)}} + \sum_{l=1}^{N} \left[ \frac{r_p^{(l)}}{\omega-\omega_p^{(l)}} - \frac{\overline{r_p}^{(l)}}{\omega+\overline{\omega_p}^{(l)}} \right] \right) \\
    \end{aligned}
    \label{eq:truncated_MOSEM}
\end{eqnarray}
where the constant, non-resonant term $\mathcal{E}_{NR}$ is a fitting parameter which accounts for the following terms from Equation~(\ref{eq:base_SEM}):
\begin{eqnarray}
    \begin{aligned}
        \mathcal{E}_{NR} = \mathcal{E}(a) - \sum_{p} \frac{r_p}{a-\omega_p}
    \end{aligned}
\end{eqnarray}

\section{Data and performance evaluation}
The truncated SEM expression in Equation~(\ref{eq:truncated_MOSEM}) can be recast as a generalized Drude-Lorentz (GDL) model as described in the main manuscript \textit{via} a change of variables:
\begin{eqnarray}
    \begin{aligned}
        &\hat{\mathcal{E}}(\omega) =  \mathcal{E}_{NR} + \frac{\gamma_0}{\omega} - \sum_{l=1}^{M} \frac{\omega_{b,l}^2}{\omega^2 + \rmi\omega\gamma_i} - \sum_{l=1}^{N} \frac{\rmi s_{1,l}\omega\Gamma_i + s_{2,l}\omega_{0,l}^2}{(\omega^2 - \omega_{0,l}^2) + \rmi\omega\Gamma_\ell}
    \end{aligned}
    \label{eq:GDL}
\end{eqnarray}
We test the GDL form with the dielectric permittivity of $9$ materials listed in table~\ref{tab:summary_exp_data}, all of which were obtained \textit{via} the Refractive Index Database~\cite{rii}.
\begin{table}[h!]
    \centering
    \begin{tabular}{ | m{0.23\textwidth}<{\centering} | m{0.25\textwidth}<{\centering} | m{0.145\textwidth}<{\centering} | m{0.165\textwidth}<{\centering} | } 
        \hline
        \multicolumn{4}{|c|}{Experimental Refractive Indices} \\
        
        \hline
        Material & Range ($\mu$m) &  Samples & Ref.\\
        
        \hline
        Gold (Au) & 0.207 - 12.40 & 57 & \cite{babar2015}\\
        
        \hline
        Silver (Ag) & 0.018 – 2.480 & 23 & \cite{werner2009}\\
        
        \hline
        Cobalt (Co) & 0.018 – 2.480 & 23 & \cite{werner2009}\\
        
        \hline
        Chromium (Cr) & 0.262 – 2.500 & 65 & \cite{sytchkova2021}\\
        
        \hline
        Hafnium dioxide ($\text{HfO}_2$) & 0.385 – 500 & 65 & \cite{bright2012}\\
        
        \hline
        Silicon dioxide ($\text{SiO}_2$) & 0.030 - 1.500 & 65 & \cite{rodriguez-deMarcos2016}\\
        
        \hline
        Titanium dioxide ($\text{TiO}_2$) & 0.300 - 1.690 & 65 & \cite{sarkar2019}\\
        
        \hline
        Tantalum pentoxide ($\text{Ta}_2\text{O}_5$) & 0.030 - 1.500 & 65 & \cite{rodriguez-deMarcos2016}\\
        
        \hline
        Graphene & 0.193 – 1.690 & 65 & \cite{song2018}\\
        
        \hline
    \end{tabular}
    \captionsetup{justification=centering}
    \caption{Summary of the experimental data used in the article. All of them are available on the Refractive Index Database~~\cite{rii}}
    \label{tab:summary_exp_data}
\end{table}
The GDL expression $\hat{\mathcal{E}}(\omega_i)=\hat{\varepsilon}_i$ evaluated at the sampled experimental frequencies $\omega_i$ is compared to the experimental data $n^2(\omega_i)=\varepsilon_i$, at the sampled frequencies $\omega_i$ of each material. The accuracy is established through the use of the relative $L_2$ and $L_\infty$ errors defined as:
\begin{eqnarray}
    \begin{aligned}
        &\text{error}_{p}\left[\hat{\varepsilon}_i, \varepsilon_i\right] = \frac{{||\hat{\varepsilon}_i - \varepsilon_i||}_p}{{||\varepsilon_i||_p}}
    \end{aligned}
\end{eqnarray}
where $p\in\{2, \infty\}$, and the norms read as:
\begin{eqnarray}
    \begin{aligned}
        &{||A_i||}_2 = \sqrt{\sum_i |A(\omega_i)|^2} \\
        &{||A_i||}_\infty = \max_i |A(\omega_i)|
    \end{aligned}
\end{eqnarray} 

\section{Retrieving the generalized Drude-Lorentz model}

The GDL model in Equation~(\ref{eq:GDL}) is optimized \textit{via} the auto-differentiation tool provided in the PyTorch library. For each of the $M$ Drude terms, two parameters are needed: $\omega_{b,l}$ and $\gamma_i$. In the Lorentz functions, there are four: $s_{1,l}$, $\Gamma_i$, $s_{2,l}$ and $\omega_{0,l}$. Accounting for the residue at $0$ and the non-resonant term, $q=2+2M+4N$ parameters are optimized in total. Even for small values of $M$ and $N$, there are thus several degrees of freedom. 
We implement fundamental physical properties as constraints on the different parameters. We want the permittivity to tends towards the vacuum permittivity as the frequency increases. We thus change the non-resonant term. In addition, the residues of purely imaginary poles must also be purely imaginary. This affects $\gamma_0$. What is more, the Lorentz terms must be associated with complex conjugate poles. Otherwise, they become Debye relaxation terms, \textit{i.e.} purely imaginary poles and residues. To obtain complex conjugate poles, we must ensure that $\Gamma_i$ does not exceed a threshold set by $\omega_{0,l}$. These constraints lead to the following expression of the permittivity:
\begin{eqnarray}
    \begin{aligned}
        \hat{\mathcal{E}}(\omega) &=  1 + \mathcal{E}_{NR} + \rmi\frac{\gamma_0}{\omega} - \sum_{\ell=1}^{M} \frac{\omega_{b,\ell}^2}{\omega^2 + \rmi\omega|\gamma_\ell|} \\ 
        &- \sum_{\ell=1}^{N} \frac{\rmi s_{1,l} \omega\Gamma_\ell + s_{2,\ell} \left[\omega_{1,\ell}^2 + (\Gamma_\ell/2)^2\right]}{\left[\omega^2 - (\omega_{1,\ell}^2 + (\Gamma_\ell/2)^2)\right] + \rmi\omega|\Gamma_\ell|}
    \end{aligned}
    \label{eq:GDL_phys}
\end{eqnarray}
Let us stress that we actually consider the relative permittivity and thus added $1$ to $\mathcal{E}_{NR}$, and that $\omega_{1,l}$ is different from $\omega_{0,l}$, since $\omega_{0,l}=|\omega_p^{(l)}|$ is the modulus of the pole, while $\omega_{1,l}=Re[\omega_p^{(l)}]$ is the real part of the pole. The imaginary part is given by $Im[\omega_p^{(l)}] = \Gamma_l/2$. We also impose that $\Gamma_l$ remains greater than $5 \times 10^{12} rad.s^{-1}$, \textit{i.e.} the complex pair of poles cannot reach the real frequency axis and are thus always stable.

We take advantage of the auto-differentiation tools of Pytorch to use a custom loss function $L(\varepsilon_i, \hat{\varepsilon}_i, \mathcal{P})$ where $\varepsilon_i$ is the experimental relative permittivity at the frequency $\omega_i$, $\hat{\varepsilon}_i=\hat{\mathcal{E}}(\omega_i)$ is the GDL expression evaluated at $\omega_i$ and $\mathcal{P}$ is the set of $q$ parameters which are optimized. We design $L$ in such a way that the real and imaginary parts of the relative permittivity are considered separately, and we adjust to weight given to either at all frequencies in order to account as much as possible for the small variations that can be observed on either of them:
\begin{eqnarray}
    \begin{aligned}
        L(\varepsilon_i, \hat{\varepsilon}_i, \mathcal{P}) &= 
        0.5\times\text{error}_2\left[\hat{\varepsilon}_i, \varepsilon_i\right] + 
        2\times\text{error}_\infty\left[\hat{\varepsilon}_i, \varepsilon_i\right] + 
        10\norm{ \frac{\hat{\varepsilon}_i - \varepsilon_i}{\varepsilon_i} }_\infty \\
         &+ 5\left<\frac{A(\omega_i)}{\delta A(\omega_i)}\right> + 5\left<\frac{B(\omega_i)}{\delta B(\omega_i)}\right> + 5\left<\frac{C(\omega_i)}{\delta C(\omega_i)}\right>
    \end{aligned}
\end{eqnarray}
where the notation $<.>$ corresponds to the averaging operation over all the sampled frequencies. $A(\omega_i)$ is the error between the real part of the model $\hat{\varepsilon}_i$ and experimental data $\varepsilon_i$, and $\delta A(\omega_i)$ accounts for the value of the real part of experimental data at the frequency $\omega_i$. It is used to change the weight of $A(\omega_i)$. $B(\omega_i)$ and $\delta B(\omega_i)$ do the same for the imaginary part, and $C(\omega_i)$ and $\delta C(\omega_i)$ directly for the modulus:
\begin{eqnarray}
    \begin{aligned}
        A(\omega_i) &= \left|\text{Re}[\hat{\varepsilon}_i - \varepsilon_i]\right|, & \delta A(\omega_i) &= \left|\text{Re}[\varepsilon_i]\right| + .1\\
        B(\omega_i) &= \left|\text{Im}[\hat{\varepsilon}_i - \varepsilon_i]\right|, & \delta B(\omega_i) &= \left|\text{Im}[\varepsilon_i]\right| + .1\\
        C(\omega_i) &= \left|\hat{\varepsilon}_i - \varepsilon_i\right|, & \delta C(\omega_i) &= \left|\varepsilon_i\right| + .01\\
    \end{aligned}
\end{eqnarray}

Finally, using the ReLu function ($\text{ReLu}(x) = max(0,x)$), we put a constraint on the imaginary part of the permittivity so that it remains positive:
\begin{eqnarray}
    \begin{aligned}
        &L_{Im} =  \sum_{i} \text{ReLu}\left( -\text{Im}[\hat{\varepsilon}_i] + 0.01 \right)
    \end{aligned}
\end{eqnarray}
The pre-factors associated with the terms of the loss function were chosen through a trial and error approach.

\begin{figure}[b!]
    \begin{center}
        \includegraphics[width=\textwidth]{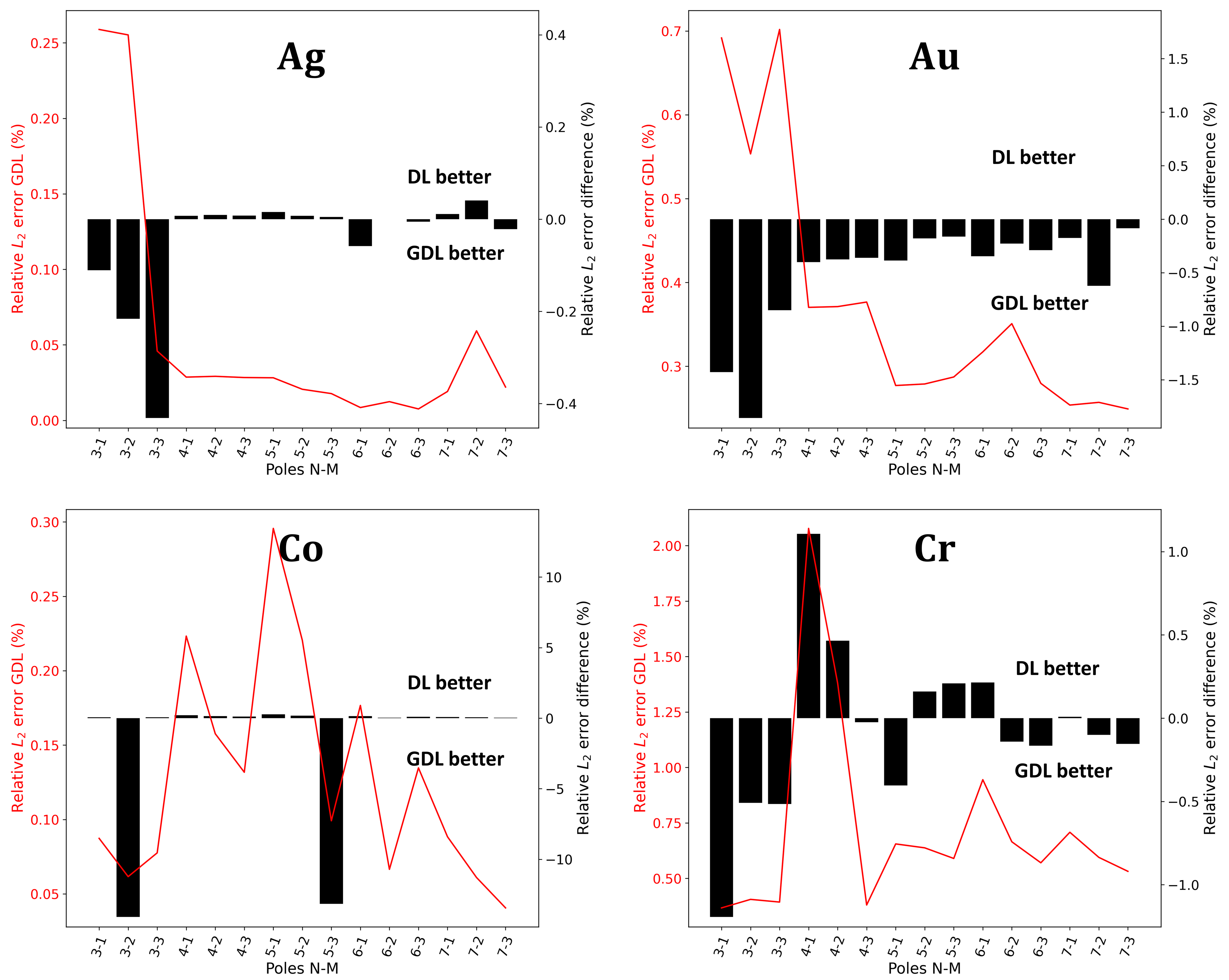}
    \end{center}
    \captionsetup{justification=centering}
    \caption{Relative $L_2$ error as a function of the number of Lorentz terms $N$ and Debye or Drude terms $M$ using the auto-differentiation approach for both the GDL model and DL model.}
    \label{Fig:figS1} 
\end{figure}

\begin{figure}[t!]
    \begin{center}
        \includegraphics[width=\textwidth]{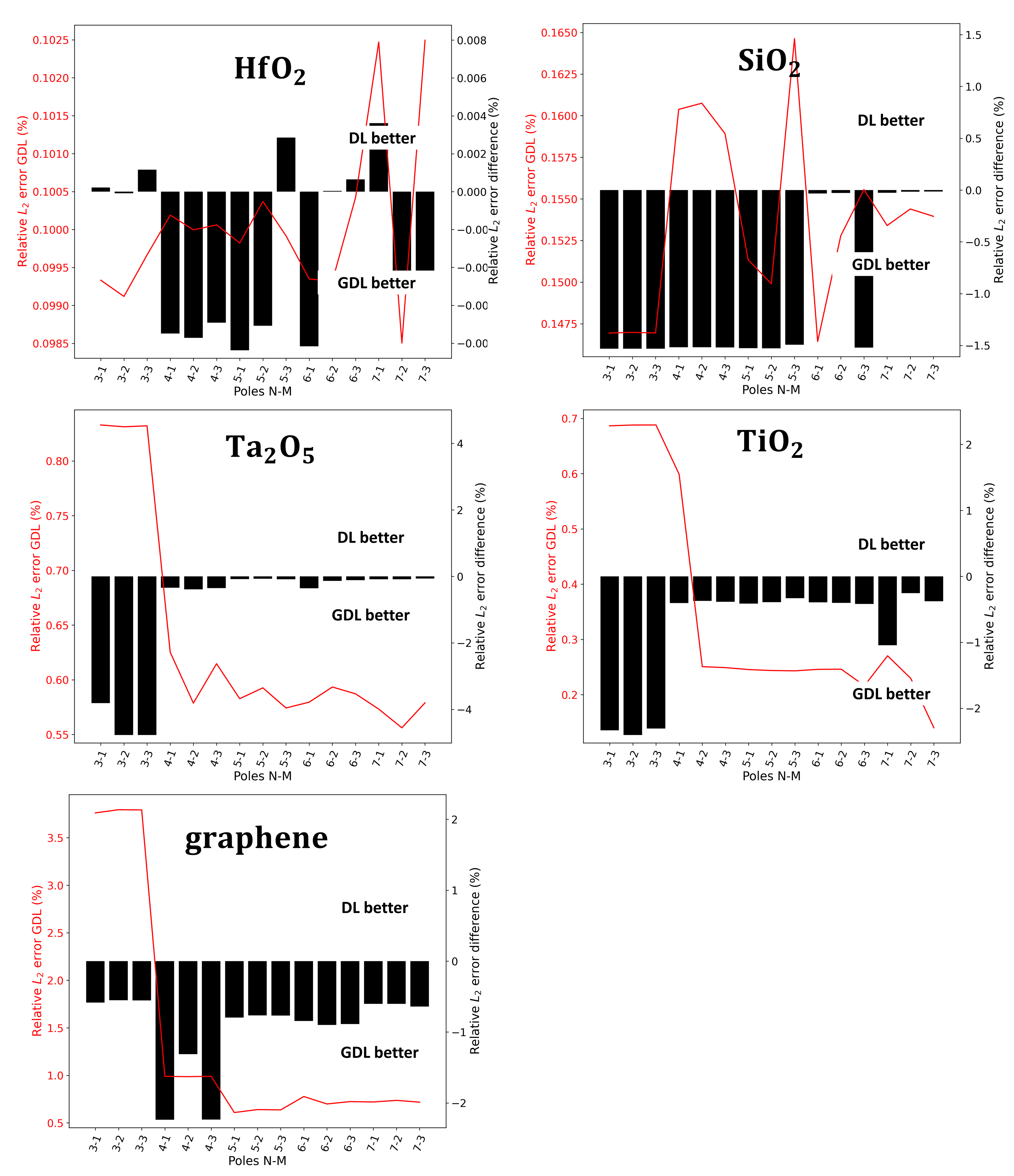}
    \end{center}
    \captionsetup{justification=centering}
    \caption{Relative $L_2$ error as a function of the number of Lorentz terms $N$ and Debye or Drude terms $M$ using the auto-differentiation approach for both the GDL model and DL model.}
    \label{Fig:figS2} 
\end{figure}

In order to select the number of Drude/Debye terms $M$ and Lorentz terms $N$, we ran the optimization process using different combinations of these parameters and calculated the relative $L_2$ error with respect to the experimental data for each material. The error curves are presented in Figures~\ref{Fig:figS1} and \ref{Fig:figS2}. The difference $\delta\text{error}_2$ between the error obtained with the GDL model and the classical Drude-Lorentz (DL) model, obtained by setting $s_{1,\ell}$ to $0$, is also shown as a bar plot. When $\delta\text{error}_2<0$, the GDL model is more accurate, which corresponds to the lower part of the plots in Figures~\ref{Fig:figS1} and \ref{Fig:figS2}.
We observe that in general, the GDL model provides more accurate expressions than the DL model when a small number of singularities are involved. This is especially true in non-metallic media. As we add more and more singularities, the models both start to account for the noise rather than the dielectric permittivity data itself, which leaves no room for comparison. In light of these observations

The numbers of terms $M$ and $N$ are then further reduced \textit{via} two means. First, we merge all the pairs of poles $\omega_a$ and $\omega_b$ which are close to one another. To do so, their barycenter is calculated using their residues $r_a$ and $r_b$ as weights, and if the barycenter provides an accurate description of the contribution of the two poles, it replaces them. Let us define $f(\omega, r_p, \omega_p)$ as: 
\begin{eqnarray}
    \begin{aligned}
        f(\omega, r_p, \omega_p) = \frac{r_p}{\omega - \omega_p}
    \end{aligned}
\end{eqnarray}
and $r_c$ and $\omega_c$ as:
\begin{eqnarray}
    \begin{aligned}
        &r_c = r_a + r_b \\
        &\omega_c = \frac{r_a}{r_c}\omega_a + \frac{r_b}{r_c}\omega_c
    \end{aligned}
\end{eqnarray}
Then, if
\begin{eqnarray}
    \begin{aligned}
        \text{error}_2\left[f(\omega, r_c \omega_c), f(\omega, r_a \omega_a)+f(\omega, r_b \omega_b)\right] \leq q_1
    \end{aligned}
\end{eqnarray}
where $q_1$ is a threshold fixed at $0.03$ through trial and error, then $p_c$ and its residue $r_c$ are used instead of ($p_a$, $r_a$) and ($p_b$, $r_b$). The second step removes all the poles which do not significantly contribute to the accuracy of the model. For each pole $\omega_p$, the relative $L_2$ error of the model $\mathcal{E}(\omega)$ deprived of $f(\omega, r_p, \omega_p)$ is calculated. If the following condition holds:
\begin{eqnarray}
    \begin{aligned}
        \text{error}_2\left[\mathcal{E}(\omega)-f(\omega, r_p \omega_p), f(\omega, r_p \omega_p)\right] \leq q_2
    \end{aligned}
\end{eqnarray}
where $q_2$ is another threhsold set at $0.15$, then the pole is removed.

\section{Performance and parameters of the optimized models}
\begin{figure}[h!]
    \begin{center}
        \includegraphics[width=\textwidth]{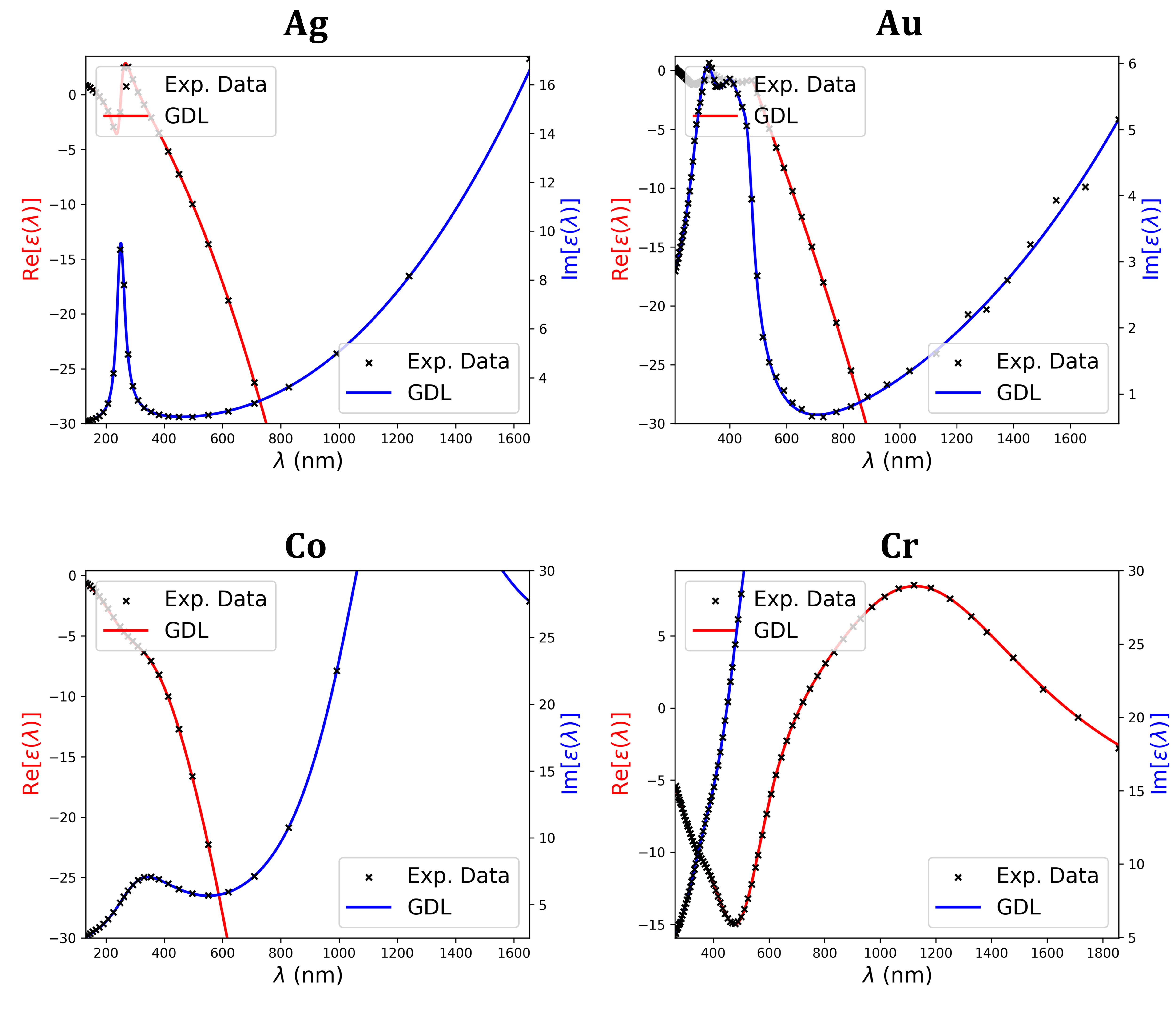}
    \end{center}
    \captionsetup{justification=centering}
    \caption{Permittivity functions calculated with the optimized GDL model compared to the experimental data.}
    \label{Fig:figS3} 
\end{figure}

\begin{figure}[h!]
    \begin{center}
        \includegraphics[width=\textwidth]{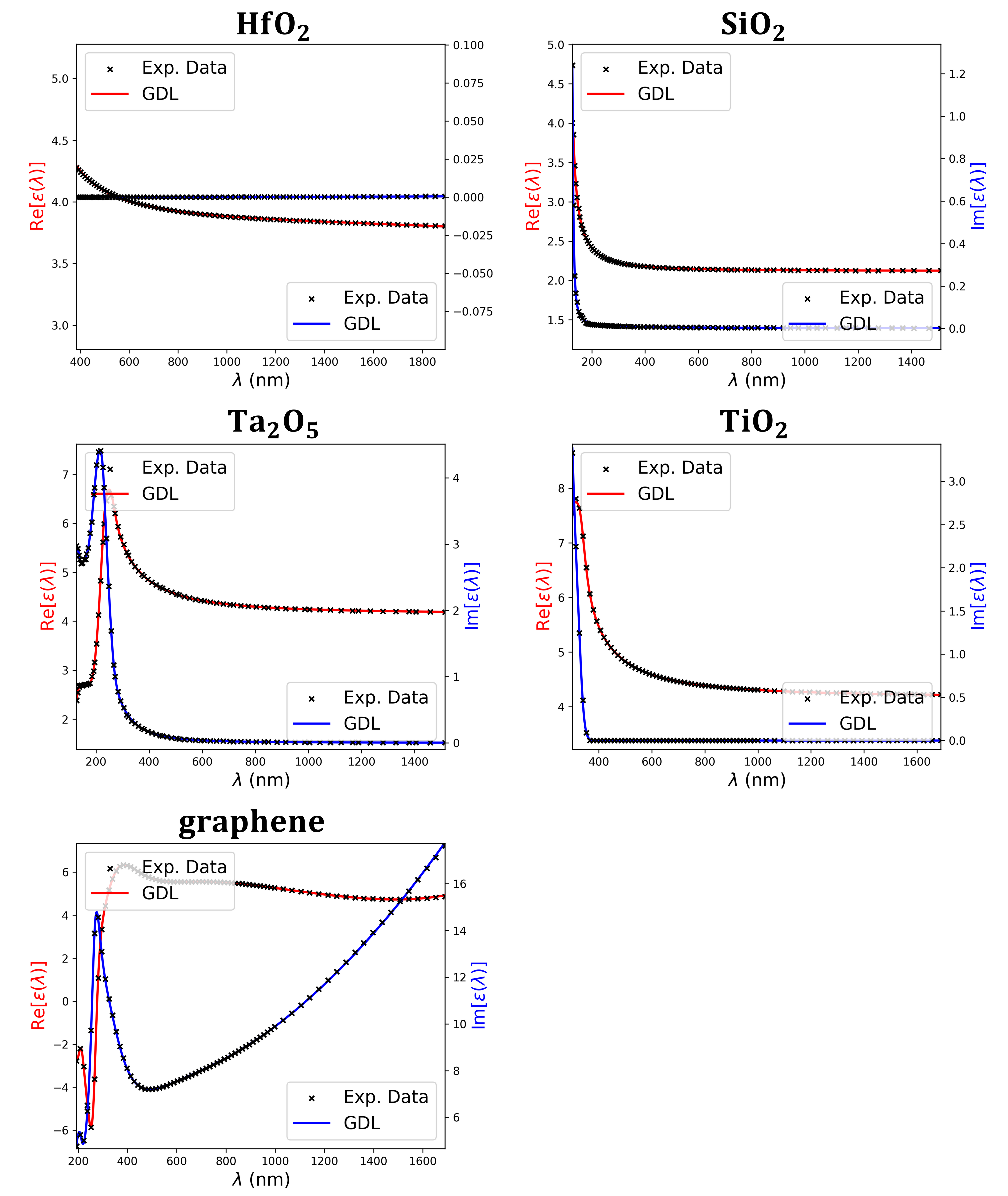}
    \end{center}
    \captionsetup{justification=centering}
    \caption{Permittivity functions calculated with the optimized GDL model compared to the experimental data.}
    \label{Fig:figS4} 
\end{figure}

The overall accuracy of the GDL model for all the materials listed in Table~\ref{tab:summary_exp_data} is presented in Table~\ref{tab:results}. The number of poles, Lorentz terms and Debye terms selected for each material, after merging and removing the irrelevant ones as described in the previous section, are specified.
\begin{table}[h!]
    \centering
    \begin{tabular}{ | m{0.17\textwidth}<{\centering} | m{0.12\textwidth}<{\centering} | m{0.12\textwidth}<{\centering} | m{0.22\textwidth}<{\centering} | m{0.22\textwidth}<{\centering} | } 
        \hline
        \multicolumn{5}{|c|}{Relative Errors of the Fitting} \\

        \hline
        Material & N & M & $\text{error}_2$ ($\%$) & $\text{error}_\infty$ error ($\%$)\\
        
        \hline
        $\text{Au}$ & 5 & 1 & 0.272 & 0.247\\
        
        \hline
        $\text{Ag}$ & 3 & 2 & 0.854 & 0.964\\
        
        \hline
        $\text{Co}$ & 5 & 1 & 0.145 & 0.070\\
        
        \hline
        $\text{Cr}$ & 3 & 1 & 0.542 & 0.834\\
        
        \hline
        $\text{HfO}_2$ & 2 & 0 & 0.074 & 0.276\\
        
        \hline
        $\text{SiO}_2$ & 2 & 0 & 0.287 & 1.192\\
        
        \hline
        $\text{TiO}_2$ & 3 & 0 & 0.237 & 0.763\\
        
        \hline
        $\text{Ta}_2\text{O}_5$ & 6 & 0 & 0.515 & 1.180\\
        
        \hline
        \text{Graphene} & 5 & 1 & 0.331 & 0.843\\
        
        \hline
    \end{tabular}
    \captionsetup{justification=centering}
    \caption{Relative errors of the optimized GDL model compared to the experimental data for the $9$ materials listed in table~\ref{tab:summary_exp_data}.}
    \label{tab:results}
\end{table}
Using the optimized expressions with the parameters from Table~\ref{tab:results}, we plot the dielectric permittivity of each material in Figures~\ref{Fig:figS3} and ~\ref{Fig:figS4}.
Finally, the list of the optimized parameters using the GDL model for all $9$ materials can be found in the Zenodo repository~\cite{zenodo2023GDL}.

\medskip
\providecommand{\noopsort}[1]{}\providecommand{\singleletter}[1]{#1}%